\documentclass[12pt]{article}
  \usepackage{amsthm,amsfonts}
  \usepackage{amsmath}
\newcommand{\bea}   {\begin{eqnarray}}
\newcommand{\eea}   {\end{eqnarray}}
\begin{document}
\date{}

\title{\textbf{Is There Flavour Independence in \\
Tensor Glueball Decays?}\footnote{Published in \textit{Europhysics Letters} \textbf{3} (6), pp.~677-680 (1987).}}

\author{F. Caruso\thanks{On leave of absence from the Instituto de F\'{\i}sica da Universidade do Estado do Rio de Janeiro, Brazil, with a fellowship from the CNPq-Brazil.}\, and E. Predazzi\\
Dipartimento di Fisica Teorica dell'Universit\`{a} di Torino -- Torino, Italy \\
Istituto Nazionale di Fisica Nucleare Sezione di Torino\\
(received 15 September 1986; accepted 3 December 1986)}

\maketitle

\begin{abstract}
The flavour independence hypothesis for (tensor) glueball decays into exclusive final states is not ruled out by the existing data. A new methodology for testing its validity and accuracy is proposed in the framework of a particular mixing scheme. As an important consequence of our analysis, a relative factor of the order of 10 must be experimentally observed for the ratio $\Gamma (\Theta \rightarrow K\bar{K})\Gamma (\Theta \rightarrow \pi \pi )$, if flavour independence indeed is valid.
\end{abstract}

It is well known that the validity of flavour independence of glueball decays into exclusive hadronic channels is still an open question in QCD phenomenology. It has sometimes been assumed, nevertheless, that this is an intrinsic property of glueballs, without a real justification. It is important, however, to stress that the phenomenological validity of this property at the hadronic level may not necessarily follow from the basic flavour symmetry at the level of quark-gluons interaction. Actually, even though this should ideally be the case, it is a conjecture that cannot yet be proved or disproved from first principles, due to our present almost complete ignorance of the process of hadronization. One should hence resort to ``conventional''  phenomenology to shed light on this fundamental subject.

The main purpose of this letter is to propose a methodology for discussing this problem in the framework of a particular mixing scheme for tensor mesons. Once we have computed all the components of this mixing scheme, it will be shown how the available experimental data for the (mixed) tensor mesons $\rightarrow \pi \pi^-$ and $\rightarrow K\bar{K}^-$ decay modes could be used to test the validity and the accuracy of the flavour symmetry hypothesis for glueball decays into exclusive hadronic final states.

In a forthcoming paper \cite{um} it is shown that no phenomenological evidence, capable of proving or ruling out, in an unambiguous way, the possibility of gluonic admixtures in the $f$-$f^{\prime}$ system, exists at present. It is our contention that the standard mixing between $f$ and $f^{\prime}$ may be considerably affected by the existence of a tensor glueball in the same mass region of these two mesons, even though a strong evidence for introducing a gluonic component into the $f$-$f^{\prime}$ system cannot be claimed by the present experimental data. The existence of such a pure gluonic state is predicted by several theoretical calculations \cite{dois}. Thus, following the same procedure of two recent papers \cite{tres}  analysing the $\eta$-$\eta^{\prime}$-$\iota$ sector, we have investigated the consequences of having a more general mixing scheme among $f$-$f^{\prime}$ and a third meson (tentatively the $\Theta$-meson) \cite{um}. Our procedure can be summarized as follows.

We start from the ansatz
\begin{equation}
|T\rangle =X_T|q\bar{q}\rangle +Y_T|\bar{S}\rangle +Z_r|gg\rangle
\label{eq1}
\end{equation}
to describe the physical states $T=f$, $f^{\prime}$, $\Theta$, where $|q\bar{q}\rangle$ is the ``ideal'' octet-singlet mixture $|u\bar{u}+d\bar{d})\sqrt{2}$, $|s\bar{s}\rangle$ is the strange-quark component and $|g\bar{g}\rangle$ the gluonic one. They are assumed to be orthogonal states. The motivation for proposing this mixing scheme is not new \cite{quatro}, but the philosophy we have followed in treating this subject is certainly a new one. All the $X_T$, $Y_T$ and $Z_T$ coefficients are deterrnined by the following steps:

i) The two-photon decay widths of the $A_2$, $f$ and $f^{\prime}$ tensor mesons are taken as input.
In particular, we use the ratios $\Gamma(f\rightarrow \gamma \gamma )/\Gamma(A_2\rightarrow \gamma \gamma)$ and $\Gamma(f^{\prime}\rightarrow \gamma \gamma)/\Gamma(A_2\rightarrow \gamma \gamma)$ \cite{cinco}. The reason is twofold. These ratios have the advantage of being independent of the gluonic amplitude, because the electromagnetic decays are due to the (``charged'') quark components. The constraints obtained on each of the components of eq.~(\ref{eq1}) are, therefore, completely independent of the flavour independence hypothesis, whose validity we are querying. Second, the $\gamma \gamma$ coupling is a most fundamental parameter of a resonance. With these two relations we are able to determine the three mixing angles as a function of the physical masses of the mesons $f$, $f^{\prime}$ and $\Theta$ and of the mass $(m_{g0})$ of the pure gluonic state. A consistent solution with real mixing angles for all the components of the mixing scheme could be obtained only if $m_{g0}$ lies into the range $1.558 \leq m_{g0} \leq 1.634$~GeV. This supports our initial guess that such a state could interfere with the $f$-$f^{\prime}$ system. By varying $m_{g0}$, we find a spectrum of ``possible'' solutions for $X_T$, $Y_T$ and $Z_T$.

ii) Other constraints on the strange and nonstrange contents follow from the results \cite{seis} on $BR \ (J/\psi \rightarrow \phi (f,f^{\prime}))$ and
$BR\ (J/\psi \rightarrow \omega (f,f^{\prime}))$, respectively. At this point the number of
possible solutions (as a function of $m_{g0}$) for $X_T$, $Y_T$ and $Z_T$ is strongly reduced.

At this point it is important to stress that a better knowledge of all the $T \rightarrow \gamma \gamma$ and $J/\psi \rightarrow VT$ decays (with $V=\omega ,\phi$) could make these two above steps sufficient enough to give a unique solution for the set $X_T$, $Y_T$ and $Z_T$ and, consequently, also for $m_{g0}$. In this case there would be no more free parameters in the model and the flavour independence hypothesis could be directly verified without ambiguities.

As the situation is not such as to allow us to use only the above two steps, to proceed further we calculate the ratios $\Gamma(f^{\prime}\rightarrow K\bar{K})/\Gamma (f\rightarrow K\bar{K})$ (assuming $BR$ $(f^{\prime}\rightarrow K\bar{K})\approx~1$), $\Gamma (\Theta \rightarrow K\bar{K})/\Gamma (f\rightarrow K\bar{K})$, $\Gamma (f^{\prime}\rightarrow \pi \pi)/\Gamma (f\rightarrow \pi \pi )$ and $\Gamma (\Theta \rightarrow \pi \pi )/\Gamma (f\rightarrow \pi \pi )$ assuming $D$-wave dominance \cite{sete}.

These quantities have already been calculated using eq.~(\ref{eq1}) \cite{sete}.
The main difference of our approach is that {\it we do not assume a priori} that the $|gg\rangle$ component couples to the two-body mesonic final
state (in the present case $\pi \pi$ and $K\bar{K}$) in an $SU(3)$-invariant manner. We rather aim at verifying whether the flavour independence
hypothesis of gluonic-hadronic coupling is compatible with the present data within our mixing scheme.

Using the numerical values $\Gamma (f\rightarrow \pi \pi ) = (150 \pm  17)$ MeV \cite{seis}, $BR(f^{\prime}\rightarrow K\bar{K} )<0.06$ (cl. 95$\%$)[8], $\Gamma (f^{\prime})=(85\pm 16)$ MeV \cite{oito}, \cite{nove} (which give $\Gamma (f^{\prime}\rightarrow \pi \pi )<(5.1 \pm  1.0)$~MeV), $\Gamma (f\rightarrow K\bar{K})=(5.2\pm  0.7)$~MeV \cite{seis}  and assuming $BR(f^{\prime}\rightarrow K\bar{K})\approx 1$ we compute the values of $r_{gK\bar{K}}=\langle K\bar{K}|gg\rangle /\langle \pi \pi |q\bar{q}\rangle$ and $r_{g\pi \pi}=\langle \pi \pi |q\bar{q}\rangle/\langle \pi\pi|q\bar q\rangle$. The advantage of calculating these quantities is that deviation from the equality $|r_{gK\bar{K}}|=|r_{g \pi \pi}|$ is a direct measurement of the deviation from the flavour independence hypothesis. It turns out that only for the solutions corresponding to the values $m_{go}=1.61$ GeV and $m_{go}=1.62$ GeV it may be possible to have flavour independence. This latter value is what one expects from lattice and potential model calculations \cite{um}.

At this point we calculate all the quantities shown in eqs. (\ref{eq4}) below for both cases $m_{g0}=1.61$ and $m_{g0}=1.62$ GeV, assuming flavour independence hypothesis. The best set of predictions seems to correspond to the value $m_{go}=1.62$ GeV. This solution gives the following mixing coefficients (see eq. (\ref{eq1})):
\begin{equation}
\left(\begin{array}{l}
f\\
f^{\prime}\\
\Theta \\
\end{array}
\right) =
\left( \begin{array}{lll}
-0.988 & \ \ 0.137 & 0.068 \\
\ \ 0.148 & \ \ 0.748 & 0.647 \\
\mp 0.038 & -0.650 & 0.759\\
\end{array}
\right)
\left( \begin{array}{l}
q\bar{q}\\
s\bar{s}\\
gg\\
\end{array}
\right)
\label{eq2}
\end{equation}
The result (\ref{eq2}) may be compared with the finding of ref. \cite{dez}
\begin{equation}
\left(\begin{array}{l}
f\\
f^{\prime}\\
\Theta \\
\end{array}
\right) =
\left( \begin{array}{lll}
\ \ 0.97 & \ \ 0.11 & 0.21 \\
-0.22 & \ \ 0.77 & 0.60 \\
-0.10 & -0.63 & 0.77\\
\end{array}
\right)
\left( \begin{array}{l}
q\bar{q}\\
s\bar{s}\\
gg\\
\end{array}
\right)
\label{eq3}
\end{equation}
where one can see that the main difference is in the sign of the $q\bar{q}$ components of the $f$ and $f^{\prime}$.
In the $f$ case this difference does not matter because this is by far the dominant component ($\approx 98\%$). However, this sign difference turns out to be responsible for the prediction \cite{sete} $\Gamma (f^{\prime} \rightarrow \gamma \gamma )\approx 0$, which does not meet the experimental indication.

The prediction we get from (\ref{eq2}) plus the flavour independence hypothesis $(r_{gK\bar{K}}\approx 0.120  \approx -r_{g\pi \pi})$
are
\begin{equation}
\left\{\begin{array}{l}
\Gamma (f^{\prime}\rightarrow \pi \pi )\ \ =(1.6 \pm 0.2)~\mbox{MeV},\\
\Gamma (\Theta \rightarrow \pi \pi )\ \ =(8.1\pm 0.9)~\mbox{MeV},\\
\Gamma (\Theta \rightarrow K\bar{K})=(59.0 \pm 8.0)~\mbox{MeV},\\
\Gamma (\Theta \rightarrow \gamma \gamma )\ \ = (0.3 \pm 0;1)~\mbox{KeV}.
\end{array}
\right.
\label{eq4}
\end{equation}
The first of the above predictions is within the present experimental limit $\Gamma_{exp}(f^{\prime}\rightarrow \pi \pi )< 5.1$ MeV. Using the mean value $\langle \Gamma_{\Theta}\rangle = 150$ MeV \cite{onze}, we get BR $(\Theta \rightarrow K\bar{Kr})\approx 0.40$ and BR $(\Theta \rightarrow \pi \pi )\approx  0.054$ (for the $m_{go}=1.61$ MeV value these quantities are reduced to approximately one-half of these values). This suggests that if flavour independence indeed exists one must find other decay modes for $\Theta$, like $\rho \rho$, $K^{\ast}K^{\ast}$, 4$\pi$ etc., and the $\pi \pi$-decay is suppressed by almost a factor 10 relative to the $K\bar{K}$-decay mode of the $\Theta$. This result is against naive expectations that if $\Theta$ is a pure glueball state (a pure singlet) one must find a substantial decay rate into $\pi \pi$ \cite{doze}.

Summarizing, the data are not in disagreement with a coherent mixing scheme for $f$-$f^{\prime}$-$\Theta$ \cite{um}. We have proposed a methodology for discussing the status of the flavour independence hypothesis for glueball decays into exclusive final states, in the framework of this general mixing scheme. Unfortunately, the overall available data are scarce. What we would like to stress is the possibility of shedding light into a fundamental open question of QCD phenomenology (namely flavour independence of glueball decays) by bettering our knowledge on ``conventional'' mesons like $f$, $f^{\prime}$ and $A_2$. We hope this possibility motivates further experimental searches.

As a last remark, we would like to remember that flavour independence of glueball decays (as well as other properties of this kind of hadron) is a nonperturbative QCD problem. As stressed in the introduction of this letter, if this problem ought to be treated at the level of fundamental constituents the problem of how the hadronization process work is immediately raised: the phenomenology of exclusive final states is among the most poorly understood aspects of QCD. In our method those perturbative effects are clearly hidden in the definitions of $r_{gK\bar{K}}$ and $r_{g\pi \pi}$. We do not calculate these quantities corresponding to nonperturbative processes like $gg\rightarrow K\bar{K}$ and $gg\rightarrow \pi \pi$. We have rather tried to give a useful parametrization for those quantities which, effectively, allow us to shed light on the problem of flavour independence from a phenomenological point of view. Perhaps when a fully consistent theory of quark-gluon mixing will be available one will get a deeper understanding of this fundamental problem.

\vspace*{0.3cm}
\centerline{ * * * }
\vspace*{0.3cm}

One of us (FC) would like to thank the Conselho Nacional de Desenvolvimento Cien\-t\'{\i}fico e Tecnol\'{o}gico (CNPq) of Brazil for financial support.

\end{document}